\documentclass[prl,preprint]{revtex4}
\usepackage{graphicx} 
\usepackage{color}
\definecolor{Red}{rgb}{0,0,0}
\definecolor{Blu}{rgb}{0,0,0}
\begin{document}
\title{I.  Dissociation free energies in drug-receptor systems via non equilibrium
alchemical simulations:  theoretical framework}
\author{Piero Procacci}
\affiliation{Dipartimento di Chimica, Universit\`a di Firenze, Via
della Lastruccia 3, I-50019 Sesto Fiorentino, Italy} 

\affiliation{Centro Interdipartimentale per lo Studio delle Dinamiche Complesse(CSDC),
Via Sansone 1, I-50019 Sesto Fiorentino, Italy}
\date{\today}
\begin{abstract}
\noindent
In this contribution I critically revise the alchemical reversible
approach in the context of the statistical mechanics theory of non
covalent bonding in drug receptor systems.  I show that most of the
pitfalls and entanglements for the binding free energies evaluation in
computer simulations are rooted in the equilibrium assumption that is
implicit in the reversible method. These critical issues can be
resolved by using a non-equilibrium variant of the alchemical method
in molecular dynamics simulations, relying on the production of many
independent trajectories with a continuous dynamical evolution of an
externally driven alchemical coordinate, completing the decoupling of
the ligand in a matter of few tens of picoseconds rather than
nanoseconds.  The absolute binding free energy can be recovered from
the annihilation work distributions by applying \textcolor{Red}{an unbiased}
unidirectional free energy estimate, on the assumption that any
observed work distribution is given by a mixture of normal
distributions, whose components are identical in either direction of
the non-equilibrium process, with weights regulated by the Crooks
theorem. I finally show that the inherent reliability and accuracy of
the unidirectional estimate of the decoupling free energies, based on
the production of few hundreds of non-equilibrium independent
sub-nanoseconds unrestrained alchemical annihilation processes, is a
direct consequence of the funnel-like shape of the free energy surface
in molecular recognition. An application of the technique on a real
drug-receptor system is presented in the companion paper.
\end{abstract}
\maketitle
\section{Introduction}
The determination of the binding affinity of a ligand for a biological
receptor system is placed right at the start of the drug discovery and
development process, in a sequence of increasing capital-intensive
steps, from safety tests, lead optimization, preclinical and clinical
trials. Thanks to modern experimental and computational techniques,
the cost for screening putative ligands for a given protein target has
diminished steadily in the last decades. Regrettably, this increased
productivity in ligands screening did not translate in a corresponding
surge in the rate of approved drugs.\cite{Munos2009} It is becoming
increasingly clear that the observed decline in the R\&D productivity
in the pharmaceutical industry in the last decades, the so-called
Eroom Law,\cite{Scannell2012} is largely due to the high cost of
failures at some stage along the drug development
sequence. Paradoxically, the screening capabilities in High throughput
screening or computer-based {\it de novo} techniques, by letting many
candidates to proceed further in the drug discovery pipeline,
unavoidably produces a sharp increase in the cost of
failure.\cite{phrma} From a computational standpoint, structure based
virtual screening using molecular docking technologies is definitely
part of the problem.\cite{Lavecchia2013,Marechal2011} The reliability
of the common docking scoring functions regarding the affinity of a
ligand for a target is severely undermined by factors such as the
complete or partial neglect of protein reorganization, microsolvation
phenomena, entropic effects, ligand conformational disorder
etc.\cite{Chodera11,Deng2015} \textcolor{Red}{The simplifying
assumptions implied in molecular docking}, while
speeding up the screening process, have in general a strong negative
impact on the predictive power of the method that is often unable to
discriminate between ligands of nanomolar, micromolar or millimolar
affinity\cite{Deng2015,Marechal2011} hence producing a large number of
costly {\it false positive}.

In the last two decades, in the context of atomistic molecular
dynamics (MD) simulations with explicit solvent, various computational
techniques have been devised to compute the absolute binding free
energies with unprecedented accuracy such as the Double Decoupling
method (DDM),\cite{gilson97} Potential of Mean Force
(PMF)\cite{Roux04,Colizzi2010}, Metadynamics\cite{Laio02,Gervasio10,Biarnes2011} or
generalized ensemble approaches (GE) like the Binding Energy
Distribution Analysis (BEDAM)\cite{bedam}, the Adaptive Integration
Method\cite{Fasnacht2004}, or the Energy Driven Undocking
scheme.\cite{edu} All these methodologies bypass the sampling
limitations that are inherent to classical molecular dynamics
simulations in drug receptor systems by appropriately modifying the
interaction potential and/or by invoking geometrical restraints so as
to force the binding/unbinding event in a simulation time scale
typically in the order of the nanoseconds.\cite{Roux13,Deng2015} In
the so-called alchemical
transformations\cite{Jorgensen85,Shirts07,Jorgensen08,Roux13,Roux09,gilson97,Gallicchio2011,Hansen2014,Wang2015},
probably the most popular and widely used \cite{Shirts07,Schulten05}
of these methods, the ligand, in two distinct thermodynamic processes,
is reversibly decoupled from the environment in the bulk solvent and
in the binding site of the solvated receptor.  Reversible decoupling
is implemented by discretizing the non physical alchemical path in a
series of independent equilibrium simulations each with a different
Hamiltonian H($\lambda_i$) with the ligand-environment coupling
$\lambda_i$ parameter varying in small steps from $\lambda=1$ to
$\lambda=0$ corresponding to the fully coupled and decoupled
(gas-phase) state of the ligand, respectively.  In most of the
variants of the reversible alchemical route, a geometrical restraint,
whose spurious contribution to the binding free energy may be
eliminated {\it a posteriori}, keeps the ligand in the binding site at
intermediate values of the $\lambda$ coupling parameter. The overall
free energies for the two decoupling processes are computed by summing
up the free energies differences relative to $\lambda$-neighboring
Hamiltonians using either thermodynamic integration\cite{Kirkwood35}
(TI) or the free energy perturbation\cite{Zwanzig54} (FEP) scheme with
the Bennett acceptance ratio.\cite{Bennett76,Shirts03,Procacci13} The
absolute standard binding free energy can be finally computed as the
free energy difference between the two decoupling
process\cite{Jorgensen85} using a
correction\cite{gilson97,General2010,Procacci2015} to account for the
reversible work needed to bring the ligand volume from that imposed in
the MD simulation to that of the standard state. The alchemical
procedure can be merged with GE approaches by letting $\lambda$
hopping between neighboring $\lambda$ states so as to favor
conformational sampling of the
ligand.\cite{Chodera11,bedam,Gallicchio2011,Kaus2015}

In this contribution I critically revise the alchemical reversible
approach in the context of the statistical mechanics theory of non
covalent bonding in drug receptor systems, evidencing the strengths
and the weakness of the methodology from a computational standpoint.
For example, although the alchemical approach to the binding free
energy determination can be effectively parallelized, still, due to
unpredictable convergence problems that may emerge at the non
physical intermediate $\lambda$ states, the CPU cost per
ligand-receptor pair remains considerable,\cite{Gallicchio2011,
  Yamashita09,Roux09,Yamashita2015,Kaus2015} with a non negligible
share\cite{Roux13,Deng2015,Yamashita09} of the overall parallel
simulation time being invested in equilibration. Besides, minimizing
the free energy variance in reversible $\lambda$-hopping alchemical
simulations without degrading excessively the performances is far from
trivial.\cite{Chodera11,Shirts2015,Kaus2015} 

We then rationalize the equilibrium unrestrained alchemical
transformations, the so-called Double Annihilation method (DAM)   
by W.L. Jorgensen and C. Ravimohan\cite{Jorgensen85},   
as a limiting case of a general non equilibrium (NE) theory of
alchemical processes, specifically addressing some controversial and
elusive issues like the volume dependence of the decoupling free
energy of the bound state.\cite{General2010,Yamashita09,pande06} We
further show that most of pitfalls and entanglements in the equilibrium
approach can be resolved by using the recently proposed non-equilibrium
variant of the alchemical method, named Fast Switching Double
Annihilation Method (FS-DAM)\cite{fsdam} relying on the production of
many independent non-equilibrium trajectories with a {\it continuous}
dynamical evolution of an externally driven alchemical
coordinate,\cite{Procacci14} completing the alchemical decoupling of
the ligand in a matter of few tens of picoseconds rather than
nanoseconds.  The absolute binding free energy is recovered from the
annihilation work distributions by applying an {\it unidirectional} free
energy estimate, on the assumption that any observed work distribution
is given by a mixture of Gaussian distributions,\cite{Procacci2015}
whose normal components are identical in either direction of the
non-equilibrium process, with weights regulated by the Crooks
theorem.\cite{Crooks98} In FS-DAM, the sampling issue at intermediate
$\lambda$ state is eliminated altogether. The accuracy in FS-DAM free
energy computation relies on the correct sampling of the initial fully
coupled state alone and on the resolution of the work distribution
depending on the number of independent NE
trajectories. With this regard, I show that the reliability and
accuracy of the unidirectional estimate of the decoupling free
energies, based on the production of few hundreds of NE independent
sub-nanoseconds unrestrained alchemical annihilation processes, is a direct
consequence of the funnel-like shape of the free energy surface in 
molecular recognition.

\section{The statistical-thermodynamic basis for non covalent binding}
The statistical mechanics foundation for the non covalent binding
in drug receptor systems in solution is based on the assumption that in the 
following chemical equilibrium
\begin{equation}
{\rm R + L \rightleftharpoons  RL }
\label{eq:eq}
\end{equation}
the complex, RL, behaves as distinct chemical species\cite{Gilson2004}
with its own chemical potential, exactly as the well defined chemical
species ${\rm R}$ and ${\rm L}$.  Because of the intrinsic weakness of
the non bonded interactions (from few to few tens of $k_B T$), the
partition function of the complex R-L must rely on the definition of
the configurational quantity $I({\bf r},\Omega)$, with ${\bf r}$ and
$\Omega$ being the translational and orientational coordinates of the
ligand relative to the receptor. $I({\bf r},\Omega)$ is equal to 1 where the complex
is formed and 0 otherwise.\cite{gilson97,luo02,Gilson2004,Zhou09} In
the infinite dilution limit, the equilibrium constant for the reaction
of Eq. \ref{eq:eq}, $K=\frac{[RL]}{[R][L]}$, can be defined
in terms of the canonical statistical average $ \langle I({\bf r},
\Omega )\rangle V |_{\lim V \rightarrow \infty}$.\cite{luo02} The
quantity $\langle I({\bf r},\Omega) \rangle $ tends to zero at
infinite dilution, such that the product $\langle I({\bf
  r},\Omega)\rangle V$ tends to the equilibrium constant as $V$ tends
to infinity:
\begin{eqnarray}
K  \equiv \langle I({\bf r}, \Omega ) \rangle V |_{\lim \rightarrow
  \infty} & =  & V
\frac{\int I({\bf r}, \Omega) e^{-\beta   w({\bf r},\Omega)} d {\bf r}
  d \Omega}
{
\int e^{-\beta w({\bf r},\Omega)} 
d {\bf r} d \Omega 
}  \nonumber 
\\
& = &   
\frac{1}{8\pi^2}
\int I({\bf r}, \Omega) e^{-\beta
  w({\bf r},\Omega)} d {\bf r} d\Omega  \label{eq:keq}
\end{eqnarray}
where $w({\bf r},\Omega)$ is the potential of mean force (PMF) for
the $\{{\bf r},\Omega \}$ ligand-receptor conformation.  In deriving
the last equation, we have used the fact that $ \lim_{V\rightarrow
  \infty} \int e^{-\beta w({\bf
    r},\Omega)} d {\bf r} d \Omega =  8\pi^2 V$, as the PMF $w({\bf
  r},\Omega)$ is non zero only in a limited volume where the RL
complex exists and zero otherwise. Eq. \ref{eq:keq} is sometimes
written as an integral restricted to the so-called {\it binding site
  volume} ${V_{\rm site}}$
 \begin{equation}
\frac{1}{8\pi^2}
\int_ {V_{\rm site}} e^{-\beta w({\bf r},\Omega)} d {\bf r} d\Omega.
\label{eq:vsite}
\end{equation}
The equilibrium constant $K$ for the reaction ${\rm R + L
  \rightleftharpoons RL }$ has the dimension of a volume and is a true
physical observable, usually accessed by measuring some spectroscopic signal $s$
that is proportional to the fraction of bound receptors (binding
isotherm). \cite{Gilson2004} The {\it binding free energy} is related
to $K$ via the equation
\begin{equation}
\Delta G = -k_B T \ln (K/V_{\rm ref} )
\label{eq:dg0}
\end{equation}
where $V_{\rm ref}$ is the reference volume in units consistent with
the units of concentration in $K$, e.g., 1 M or about 1661
$A^3$/molecule for molarity units. As such, the free energy defined in
Eq. \ref{eq:dg0} is a purely {\it conventional} quantity, measured
with respect to some state defined by the reference
molecular volume $V_{\rm ref}$. When the reference concentration is
taken to be 1M (or, equivalently, the molecular volume is 1661
\AA$^3$), $\Delta G$ corresponds to the {\it standard} binding free
energy, indicated with $\Delta G_{0}$.

In atomistic molecular dynamics simulations, the equilibrium constant
can be directly accessed by means of Eqs. \ref{eq:keq} and \ref{eq:vsite}
using PMF-based technologies\cite{baron10,Roux09} or binding energy
distribution methods.\cite{bedam} These techniques require a prior
knowledge of the domain where $I({\bf r}, \Omega)=1$. However, if the
binding is tight, and if the domain is chosen large enough so as to
include all states contributing significantly to the integral of
Eq. \ref{eq:vsite}, then the equilibrium constant is independent,
within certain limits, on the integration
domain.\cite{gilson97,bedam} Alternatively, one can compute the
free energy gain/loss in the formation/dissociation of the complex RL
starting from the unbound state in solution or {\it
  viceversa}.\cite{Roux09,Yamashita2015,Wang2015} In reversible
alchemical transformations, as we shall see later on in detail, the
free energy cost for bringing the ligand from the bound to the
unbound state in solution is obtained by constructing a thermodynamic
cycle whereby the ligand, in two distinct thermodynamic processes,
is reversibly decoupled (i.e. brought to the gas-phase) from the
environment in the bulk and in the binding site. While the decoupling
free energy of the ligand in the bulk, $\Delta G_{\rm L}$, bears no
dependence on the reference state, \textcolor{Red}{when alchemically decoupling the
ligand in the complex, the computed
free energy $\Delta G_{\rm RL}$ depends on the effective reference
concentration of (or volume available to) the ligand implied in the
simulation.}\cite{gilson97,karplus03,Zhou09,General2010} For example,
when the RL complex is {\it unrestrained} except for periodic boundary
conditions, the volume available to the ligand is apparently that of
the simulation box\cite{karplus03,Roux09,General2010,Procacci2015}.
Alternatively, one could allow the ligand in the bound state to move
within an effective volume set by a translational and rotational
restraint potential\cite{gilson97} possibly matching the region where
the function $I({\bf r}, \Omega)$ is equal to 1.  Whatever the approach
adopted, in order to make the computed dissociation free energy $\Delta
G_{\rm sim}=\Delta G_{\rm RL}-\Delta G_{\rm L} $ independent of the simulation
conditions, a standard state correction (SSC) must be added such that
\begin{equation} \Delta G_0 =
  \Delta G_{\rm sim} + k_BT \ln \left ( \frac{V_{\rm ref}} {V_{0}} \right ) + k_BT \ln \left ( \frac{\xi_{\rm ref}} {8\pi^2} \right ).
\label{eq:ssc}
\end{equation}
The second and third terms in Eq. \ref{eq:ssc} may be viewed as the
reversible work to bring the volume and the solid angle available to the
ligand in the simulation of the bound state to that of the standard
state $V_{0}=1661$ \AA$^3$ and $\xi_0=8\pi^2$,
respectively.\cite{gilson97} Eq. \ref{eq:ssc} is valid, provided the
alchemical transformation is done reversibly, that is, each
intermediate state along the alchemical decoupling coordinate {\it
  must be at equilibrium}, sampling canonically all the configurations
of the ligand contributing to the integral of Eq. \ref{eq:keq}.  In
the unrestrained alchemical approach ($V_{\rm ref}=V_{\rm box}$ and
$\xi_{\rm ref} = 8 \pi^2)$), full canonical sampling at small
$\lambda$ is pathologically difficult,\cite{karplus03} but also in
the constrained variant, the restraint can be unintentionally
implemented in a such a way that some important orientations
contributing to $I_i({\bf r}, \Omega)$ are rarely accessible or
poorly sampled in the time of the simulation.  Thus, the lack of
dependence of $\Delta G_{\rm sim}$ on ${V_{\rm ref}}$, sometimes
observed in reversible alchemical simulations, indicates a problem,
typically a convergence issue, such as the ligand not sampling the
full available phase space.

The elementary theory sketched out above works very well if the ligand
behaves as an entity performing small librations in a regular and
smooth potential set by the surrounding receptor.  In the real world
of drug-receptor binding processes, the potential in the binding site
can be very rugged, characterized by many local energy minima along
complex ro-vibrational collective coordinates.  Energetically distinct
conformations are very challenging in equilibrium based MD techniques,
\textcolor{Red}{as the final result may depend on the chosen initial
  set up of the simulation}.\cite{bedam,Kaus2015} In the simple
language of docking, we say that the ligand can adopt different
possible conformational ``poses'' with different scoring
functions. Let's now assume that the ligand can occupy the binding
site region, the so-called ``exclusion zone'' in the
receptor\cite{Gilson2004}, with different orientations. We can hence
define {\it non overlapping} step functions of the kind $I_i({\bf r},
\Omega)$ (where the index $i$ label the (RL)$_{i}$ orientational pose)
in such a way that $I({\bf r},\Omega)=\sum_i^{N_p}I_i({\bf
  r},\Omega)$, with $N_p$ being the number of poses.  In this manner,
the equilibrium concentration of the bound species RL detected by the
signal $s$ (that is assumed to be unable to discern orientational
poses in the exclusion zone) is given by
\begin{equation}
[{\rm RL}]=\sum_i^{N_p} [{\rm (RL)}_i]
\label {eq:poses} 
\end{equation} 
The species (RL)$_i$, each defined by its own $I_i({\bf r}, \Omega)$
function, are subject to the simultaneous equilibria
\begin{eqnarray}
{\rm R + L} & \rightleftharpoons  & {\rm (RL)}_i ~~~~ i=1,2,...N_p \nonumber \\
{\rm R + L} & \rightleftharpoons  & { \rm RL}
\label{eq:meq}
\end{eqnarray}
From Eq. \ref{eq:poses} we trivially obtain that the overall
equilibrium constant for the reaction ${\rm R + L \rightleftharpoons
  RL }$ can be written as $K=\sum_i^{N_p} K_i$, with $K_i =
\frac{[{\rm (RL)}_i]}{[R][L]}$ being the equilibrium constant for the
complex in the $i$-th pose.  Using Eq. \ref{eq:dg0},  we may thus define
the standard binding free energy for pose $i$ as $\Delta G_{0i} =
-k_B T \ln K_i $ where $K_i$ is expressed in molarity and a
standard state concentration of 1M is implied. Now, the molecular recognition
machinery in biological system works well because very often one
particular pose is preferred with respect to all others. If we set
$\Delta G_{01}$ as the most stable pose among the $N_p$ possible
ligand bound states, then, using Eqs. \ref{eq:dg0} and \ref{eq:poses}
we can write
\begin{eqnarray}
e^{-\beta \Delta G_0} & = & e^{-\beta \Delta G_{10}} ( 1+ \sum_{i=2}^{N_p} e^{-\beta \Delta
  G_{1}^{i} }) 
\label{eq:onepose1} 
\end{eqnarray}
where we have defined the {\it relative} free energy difference
$\Delta G^{i}_{1}=\Delta G_{0i} - \Delta G_{01},~ i=2..N_{p}$ between
pose $i$ and the most stable pose ($i=1$).  Note that the {\it
  positive} quantity $\Delta G^{i}_{1}\equiv -k_BT \ln
\frac{K_i}{K_1}$, referring to a process involving no changes in the
number of species, bears no dependence on the standard state. If all
these relative free energies $\Delta G^{i}_{1}$ are worth several
$k_BT$ such that $\sum_{i=2}^{N_p} e^{-\beta \Delta G_{1}^{i} } \ll
1$, then taking the logarithm of Eq. \ref{eq:onepose1} one may write
\begin{equation}
\Delta G_{0} = \textcolor{Red}{ \Delta G_{10} - k_B T \sum_{i=2}^{N_p} e^{-\beta \Delta
  G_{1}^{i} } }   \simeq \Delta G_{10} 
\label{eq:onepose2}
\end{equation}
where we have used the fact that $\log(1+x) \simeq x$ for $x$ small.
Eq. \ref{eq:onepose2} says that, if one of the poses is much more
stable than all the others, then the overall standard binding free
energy in the drug-receptor system is dominated by that of the most
favorable pose. Eq. \ref{eq:onepose2} is indeed at the very heart of
molecular recognition in biological systems.  Eq. \ref{eq:onepose2} is
also central,  as we shall see in the following, in the NE
theory of alchemical transformations since when it holds, a very simple
and unbiased estimate of $\Delta G_{10}$  may be derived from the work
distributions obtained in the NE trajectories.  

\section{Reversible alchemical transformations in drug receptor systems} 
As previously outlined, in the alchemical method, the absolute
standard {\it dissociation} free energy for the reaction ${\rm RL
  \rightleftharpoons R + L }$ may be recovered as the difference
between the decoupling free energy of the ligand in the binding site
and in the bulk solvent.\cite{Jorgensen85} In either processes, the
free energy along the non physical path between the fully coupled
state and the decoupled states, with Hamiltonians
$H(x,\lambda)_{\lambda=1}$ and $H(x,\lambda)_{\lambda=0}$
respectively, is computed by discretizing the $\lambda$ parameter in a
number of $N_{\lambda}$ intermediate states $\lambda_i$ in the [0,1]
interval and by running a standard MD simulation for each of these
states.  The switching off protocol of the ligand-environment
interactions may vary from system to system although there is a
general consensus for first turning off the electrostatic interactions
followed by the Lennard-Jones atom-atom terms supplemented with a
soft-core potential\cite{Beutler99,Buelens2012} to avoid catastrophic numerical
instabilities when approaching to $\lambda=0$.  The {\it reversible}
work of the whole process can be obtained by appropriately summing up
the individual free energy differences between neighboring $\lambda_i$
states evaluated using either TI or FEP techniques. In FEP, these
differences are computed exploiting the superposition of neighboring
potential energy distribution functions and implementing the Zwanzig
formula\cite{Zwanzig54} such that $\Delta G_f = -k_B T
\sum_{i=1}^{N-1} \langle e^{-\beta(H_{\lambda_{i+1}}-H_{\lambda_{i}})}
\rangle_{\lambda_{i}}$ with $\lambda_1=1$ and $\lambda_N=0$.  In the
decoupling process of the {\it bound state} RL, the ligand, for each
$\lambda_i$ state, must sample all the attainable conformations for
the given Hamiltonian $H(x,\lambda_i)$, including all secondary poses
of the kind ${\rm (RL)}_j,~j=2,3..$ (see
Eqs. \ref{eq:poses}-\ref{eq:onepose2}).  When $\lambda_i$ is
approaching to zero, \textcolor{Red}{the ligand may occasionally leave
  the receptor, severely slowing down the
  convergence.}\cite{karplus03} Therefore, when the ligand is decoupled
in the bound complex, usually it is common practice to impose a
geometrical restraint in the simulations so as to avoid the
``wandering ligand problem'' related to the choice of $V_{\rm
  ref}=V_{\rm box}$.\cite{gilson97,bedam,Roux13} The free energy cost
of imposing the restraint, the so-called ``cratic'' free
energy\cite{Hermans1997}, corresponds to the SSC discussed in
Eq. \ref{eq:ssc}.  The SSC, stemming from the restraint volume $V_{\rm
  ref}$, may be evaluated analytically\cite{Hermans1997,gilson97}, or
numerically,\cite{Roux13} depending on how the restraints are imposed.
Decoupling with restraints is often referred as Double Decoupling
Method\cite{gilson97} (DDM) while the unrestrained variant is known as
double annihilation method (DAM).\cite{Jorgensen85,pande06} In modern
DDM implementation,\cite{Roux13} the translation and rotational
restraints, that force the ligand to explore a restricted
orientational and positional space in the binding region, are
progressively enforced/removed while the ligand is being
decoupled/coupled. Hence, each $\lambda_i$ point in the [0,1] interval
is actually characterized by a potential coupling parameter
$\lambda_i^C$ and by a restraint $\lambda_i^R$ state. If each of the
$\lambda_i$ independent simulations in the [0,1] interval has reached
convergence, \textcolor{Red}{canonically sampling all conformations
  that are attainable at the Hamiltonian
  $H(x,\lambda_i^C,\lambda_i^R)$), then the free energy computed in
  either directions (decoupling or coupling) must be identical and
  independent of the initial set up of the system}.  When applying FEP
  to reversible alchemical transformations is common
  practice\cite{Yamashita09,Pohorille2010,Roux13} to evaluate the free
  energy difference between neighboring states using {\it
    bidirectional estimator}.\cite{Shirts03,Bennett76} One can in fact
  define a ``reverse'' free energy estimate as $\Delta G_r = k_B T
  \sum_{i=2}^{N} \langle e^{-\beta(H_{\lambda_{i-1}}-H_{\lambda_{i}})}
  \rangle_{\lambda_{i}}$ that must coincide for each $\lambda_i$ point
  with the forward estimate $\Delta G_f$ if equilibrium is reached
  everywhere along the alchemical coordinate. The forward and reverse
  estimate in the $\lambda$ interval (0,1) can be combined using the
  Crooks theorem\cite{Crooks98} and the Bennett acceptance
  ratio.\cite{Bennett76} The manifestation of a hysteresis is usually
  syntomatic of lack of complete convergence. The latter is often
  related to the presence of secondary poses $ {\rm (RL)}_i$ that may
  emerge especially at small $\lambda$ values,\cite{bedam,Kaus2015}
  when most of the ligand-environment interaction has been switched
  off and barriers between alternate ligand conformations/poses are
  smoothed. Kinetic traps provided by alternate poses may degrade the
  overlap between energy distributions of neighboring $\lambda_i$
  states, making the convergence slow and uneven in the [0,1]
  interval.\cite{Shirts2015} To overcome this serious problem and
  unpredictable behavior, in a parallel environment, alchemical
  transformations can be coupled to Generalized Ensemble techniques
  whereby each replica of the system performs a random walk in the
  $\lambda$ domain with $\lambda$ moving according to a Metropolis
  criterion, so as to make the $\lambda$ probability distribution flat
  on the whole [0,1] $\lambda$ interval. These methods are termed
  $\lambda$-hopping schemes and use either Hamiltonian Replica
  Exchange (HREM)\cite{Gallicchio2011}, Serial Generalized Ensemble
  (SGE) methodologies\cite{Chelli10} or Adaptive Integration Schemes
  (AIM) \cite{Fasnacht2004,Kaus2015} and are all aimed at defeating
  the convergence problems induced by the existence of meta-stable
  conformational states of the bound ligand along the alchemical path.
  \textcolor{Red}{In the HREM implementation, no bias potential is
    needed in the transition probability, while in SGE or AIM, the
    bias potential (i.e. an estimate of the free energy difference
    between neighboring $\lambda$ windows) is evaluated on the fly
    using the past history produced by all replicas.\cite{Chelli10}}

\textcolor{Red}{When different poses of the ligand in the binding site
  are separated by energy barrier significantly higher than $k_B T$,
  or for bulky ligands characterized by a manifold of conformational
  states, $\lambda$-hopping schemes may be non
    resolutive, still being plagued by convergence
    issues.\cite{Kaus2015}} For example, for a ligand as simple as
    phenol in Lysozime, convergence of the decoupling free energy
    starting form a random pose may take as much as one nanosecond of
    parallel simulation, even adopting a very fine grid when
    approaching to the decoupled state $\lambda=0$. \cite{bedam} In
    the Thrombin-CDB complex\cite{Kaus2015} after about five
    nanoseconds of $\lambda$-hopping simulation, convergence is not
    even in sight.\cite{Kaus2015} There are finally some pathological
    examples where even $\lambda$-hopping schemes exhibit a marked initial
    pose dependence, like in the \textcolor{Red}{BACE1
    complexes.\cite{Cumming2012,Kaus2015} The relative free energy of the
    BACE-24 and BACE1/17a systems may differ by as much as 4 kcal mol$^{-1}$
    with two possible symmetrical orientation of a phenyl ring of
    ligand 24 bearing a bulky substituent,} whose size makes virtually
    impossible the flipping of the ring in the binding site. In that
    case, even with the use of soft-core potentials, no mixing
    whatsoever of the two poses at any $\lambda$ state can be
    observed. One obvious way of circumventing the lack of mixing in
    these cases is of course that of increasing the density of
    $\lambda$ states near the critical points of the $\lambda$ path,
    correspondingly increasing the number of replica and the cost of
    the simulation.  Alternatively, as proposed in
    Ref. \cite{Kaus2015}, one can supplement the $\lambda$-hopping
    method with {\it ad-hoc} Hamiltonian scaling schemes on
    appropriate collective/conformational coordinates of the
    ligand. These latter approaches, however, while preserving the
    efficiency of the alchemical calculation, are not general as they
    require prior knowledge of the topology of the barriers and of the
    kinetic traps preventing the mixing between the competing poses.

Summarizing,  we may state that the real problem in reversible
alchemical simulations is related to the fact that it is not yet
available a universal protocol for minimizing the statistical
uncertainty of calculations performed along an alchemical path.
Uncertainty may depends critically on the specific subsets of
the $\lambda$ path where activated collective coordinates, possibly
induced by the imposed restraints, can cause the insurgence of kinetic
traps degrading the energy overlap of neighboring $\lambda$
states. With this regard, it has been pointed out that minimizing the
overall statistical uncertainty is equivalent to minimizing the
thermodynamic length, that is,  of choosing the $\lambda$ alchemical
protocol so that the total uncertainty for the transformation is the
one which has an equal contribution to the uncertainty across every
point along the alchemical path.\cite{Shirts2015} The quest for the
optimal path in alchemical reversible transformations is intimately
connected to the necessity of having an {\it a priori} estimate of the
accuracy in binding free energy evaluation.  The latter is indeed an
essential requirement in the development of a second generation high
throughput virtual screening tool in drug discovery.\cite{Wang2015}
In the present stage, in spite of the many noticeable efforts in this
direction, reversible alchemical transformations are still quite far
form being that tool.\cite{Shirts2015,Chodera11,Kaus2015} In the
following sections I shall discuss some aspects of the theory of non
covalent binding in the context of non equilibrium transformation,
showing that fast-switching alchemical simulations\cite{fsdam} may
provide a reliable and efficient instrument in drug discovery.

\section{Non-equilibrium theory of alchemical transformations in non covalent binding}

\subsubsection{Basic theory}
The requirement of an equilibrium transformation along the entire
(0,1] semi-open interval is lifted altogether in the recently
proposed Fast Switching Double Annihilation
Method.\cite{fsdam,Procacci2015} FS-DAM implies an equilibrium
sampling {\it only on one extreme} of $\lambda$ [0,1] interval,
i.e. at the fully coupled states of the complex and of the free ligand
in solution at $\lambda=1$. Once the initial states have been somehow
prepared, several fast non equilibrium trajectories ($N_{\tau}$) are
launched in parallel with zero communication overhead by switching off
the ligand-environment interactions in a protocol. \textcolor{Red}{The
fast decoupling protocol,  identical for all
trajectories, is analogous to that used in the reversible
counterpart} (i.e. we first switch off the electrostatic
interactions and then we turn off the dispersive-repulsive term using
a soft-core potential to avoid instabilities at low $\lambda$'s).  The
duration $\tau$ of the Non Equilibrium experiments ($\tau$-NE) may
last from few tens to few hundreds of picoseconds depending on the
size of the ligand.\cite{Procacci14} The annihilation of the ligand (in
the complex or in bulk) is conventionally taken to be the {\it
  forward} process.\cite{procacci06} The non equilibrium annihilation
or forward work, $W_{1 \rightarrow 0}$, done in driven $\tau$-NE
experiments starting from canonically sampled fully coupled states
with a common time schedule, obeys the Jarzynski theorem\cite{jarzynski97}
\begin{equation}
e^{-\beta \Delta G_{1 \rightarrow 0} } = \langle e^{-\beta W_{1
    \rightarrow 0} } \rangle  = \int P(W_{1 \rightarrow 0}) e^{-\beta W_{1
    \rightarrow 0} } dW_{1 \rightarrow 0}
\label{eq:jarzy}
\end{equation}
with $\Delta G_{1 \rightarrow 0}$ being the annihilation/forward free
energy.  For the annihilation of the complex, the free energy must
include a standard state correction that I shall discuss in detail
below in this section.  The Jarzynski formula, Eq. \ref{eq:jarzy}, is
of little practical use for evaluating $\Delta G_{1 \rightarrow 0}$
since it relies on an exponential average over the distribution
$P(W_{1 \rightarrow 0})$ on its left tail, i.e. a statistics that is
both inherently noisy and biased, even if the spread of the work data
is only moderately larger than
$k_BT$.\cite{hummer01,Gore2003,park04,Pande05,Oberhofer05} In
case of Gaussian work distributions for the (forward) annihilation
process, the Crooks theorem\cite{Crooks98},
\begin{equation}
\frac{P_{AB}(W_{1 \rightarrow 0})}{P_{BA}(-W_{0 \rightarrow 1})} =
e^{\beta(W_{1 \rightarrow 0}-\Delta G_{1 \rightarrow 0})}, 
\label{eq:ct}
\end{equation}
imposes that the underlying reverse work distribution $P_{BA}(-W_{0
  \rightarrow 1})$ for the fast growth process must also be Gaussian
with the same variance $\sigma$ and with mean work given by $\langle
-W_{0 \rightarrow 1} \rangle= \langle W_{1 \rightarrow 0} \rangle -
\beta \sigma^2$\textcolor{Red}{,\cite{hummer01,park04,procacci06,Goette2009,Gapsys2015,Procacci2015}} hence
providing an {\it unbiased unidirectional estimate} of the
annihilation free energy based on the forward process alone of the
form
\begin{equation}
\Delta G_{1 \rightarrow 0} = \langle W_{1 \rightarrow 0}  \rangle - \frac{\beta \sigma^2}{2}  
\label{eq:jarzy2}
\end{equation}
where the first two cumulants $\langle W_{1 \rightarrow 0}\rangle$ and
$\sigma$ are both a monotonic functions of duration time $\tau$ of the
NE process.\cite{Feng08,Procacci10} The term $\frac{\beta
  \sigma^2}{2}$ represents the mean dissipation during the $\tau$-NE
transformation. In this regard, it has been
observed\cite{Goette2009,Procacci14,Gapsys12} that \textcolor{Red}{the work
  distribution obtained in fast $\tau$-NE annihilation/creation
  experiments of small to moderate size organic molecules in polar
  (water) and non polar (octanol) solvents} has a marked Gaussian
character and that the corresponding dissipation is surprisingly
small.  Hydrophobic or polar molecules, annihilated/created in
explicit water or octanol in times as short as 63 or 180 picoseconds,
consistently exhibit\cite{Procacci14} dissipation energies ranging
from 1 to 2 for kcal mol$^{-1}$ for water and from 2 to 4 for kcal
mol$^{-1}$ for octanol.  \textcolor{Red}{The corresponding forward
  and reverse distributions, $P_{AB}(W_{1 \rightarrow 0}),~P_{BA}(-W_{0
    \rightarrow 1})$, have a high degree of superposition and are
  strikingly symmetrical with respect to the free energy $\Delta G_{1
    \rightarrow 0}$ in all analyzed case, as predicted by
  Eq. \ref{eq:ct} for Gaussian distributions.}  The Gaussian nature of
the annihilation/creation of small molecules in water may be
quantified by the cumulants of the distribution of order higher than
two, that according to Marcinkiewicz theorem\cite{Marcinkiewicz1939}
should all be equal to zero.  When the dissipation is small, i.e. the
spread of the distribution is limited, the Gaussian estimate
Eq. \ref{eq:jarzy2} is astonishingly robust.\cite{fsdam,Procacci2015}
\begin{table}[h!]
\begin{tabular}{c|rc|c}
\hline \hline & $\langle W_{1 \rightarrow 0} \rangle$ & $\Delta G_{1
  \rightarrow 0}$ (Eq. \ref{eq:jarzy2}) & $\Delta G_{1 \rightarrow 0}$
(Eq. \ref{eq:ct})\\ \hline Benzene & 1.75 $\pm$ 0.09 & 0.87 $\pm$0.14
& {\bf 0.79} $\pm$0.04 \\ Benzamide & 11.15 $\pm$ 0.16 & 9.95
$\pm$0.24 & {\bf 9.78} $\pm$0.07 \\ Ethanol & 4.39 $\pm$ 0.05 & 3.80
$\pm$0.04 & {\bf 3.80} $\pm$0.04 \\ Pentane & -1.59 $\pm$ 0.06 & -2.52
$\pm$0.08 & {\bf -2.56} $\pm$0.05 \\ \hline \hline
\end{tabular}
\caption{Decoupling mean work and corresponding free energies (in kcal
  mol$^{-1}$) for some polar and apolar molecules
computed  using the work distributions reported in Figure 5 of Ref. 
\cite{Procacci14}}
\label{tab:gauss}
\end{table}
in Table \ref{tab:gauss} I report results for the decoupling free
energy of drug-size molecules in water using the work data obtained in
Ref. \cite{Procacci14} for the fast switching annihilation of a set
polar and non polar molecules in water solvent in standard conditions.
The overall NE process lasted in all cases only 63 picoseconds and the
work distributions were obtained using 256 NE annihilation/growth
works.  In Table \ref{tab:gauss}, the Gaussian estimate using
Eq. \ref{eq:jarzy2} on the decoupling distribution reported in Figure
5 of Ref. \cite{Procacci14} is compared to \textcolor{Red}{the
  bidirectional estimate (in bold font) obtained by applying the
  Crooks theorem and the Bennett acceptance ratio.}  Remarkably, the
fast annihilation Gaussian estimates of the solvation free energies
are practically coincident with the maximum likelihood Bennett-Crooks
bidirectional estimate confirming the reliability of
Eq. \ref{eq:jarzy2} in fast switching alchemical transformations in
water solvent.  Regarding the errors reported in table \ref{tab:gauss}
for the Gaussian estimate, it should be remarked that
\textcolor{Red}{the variance in $\langle W\rangle$ and $\sigma^2$ for
  normally distributed samples follows the ancillary
  t-statistics\cite{wiki06} and is proportional to
  $\sigma(\tau)/(N_{\tau})^{1/2}$ and
  $\sigma^2(\tau)/(N_{\tau})^{1/2}$, respectively where $\sigma(\tau)$
  is the $\tau$-dependent spread of the underlying normal
  distribution.} So, if $\sigma$ is of the order of few kcal mol
$^{-1}$ and if Eq. \ref {eq:jarzy2} holds, {\it only few hundreds
  trajectories are needed to get an error on the free energy below 1
  kcal mol$^{-1}$}.  Unlike in reversible alchemical transformations,
in their NE variant the overall error can therefore be very naturally
and reliably computed via standard block-bootstrapping from the
collection of $N_{\tau}$ NE works. \textcolor{Red}{In Table
  \ref{tab:gauss}, for example, the errors were computed using random
  bootstrap samples with 128 work values, taken from the pool of 256
  works.} Moreover, reducing the number of NE
  trajectories by a factor of $G$ amplifies the error on $\langle
  W\rangle$ and \textcolor{Red}{$\sigma^2$} only by $G^{1/2}$ making Gaussian based
  estimates extremely robust and reliable even with a very small
  number of sampling trajectories.\cite{Procacci2015} The Gaussian
  shape in the rapid annihilation of the ligand (in the bound or in
  the unbound state) is a natural consequence of the time scale used
  in the annihilation (few tens to few hundreds of ps) of ligands in
  standard conditions. As we shall discuss in detail further below,
  such time scale is way too fast to allow extensive conformational
  sampling while $\lambda$ is continuously decreased, but is slow
  compared to the time scale of the modulating vibrational motions of
  the atoms surrounding the annihilating ligand.  In this way, the
  energy change at a given time $t$ during the driven $\tau$-NE
  process depends to a very good approximation only on the alchemical
  state (i.e. on the instantaneous value of $\lambda(t)$) at that
  given time) as in Markovian memory-less
  processes.\cite{park04,Procacci14}

\subsection{Free energy estimates for a mixture of Gaussian processes}
Eq. \ref{eq:jarzy2}, based on a single symmetrically related forward
and reverse work distributions, implies that the $\tau$-NE process
connects two well defined thermodynamic states, each defined by {\it a
  single free energy basin}. This could be the case for the process of
fast annihilating/growing of a small and relatively rigid molecule in a
solvent.  When the initial and/or the final thermodynamic states are
characterized by a manifold of free energy basins with uneven well depth,
(like for the many alternate poses of a ligand on a receptor or for
the misfolded states of a protein) then Eq. \ref{eq:jarzy2} is no
longer valid and the observed forward and reverse work distribution
can be strongly asymmetrical.\cite{procacci06} In
Ref. \cite{Procacci2015} it was shown that, in systems characterized by
a principal free energy basin and a manifold meta-stable states on one
or both end of the $\tau$-NE process, then the asymmetrical forward
$P(W_{1\rightarrow 0})$ and reverse distributions $P(-W_{0\rightarrow
  1})$ can be rationalized in terms of a mixture of an equal number
of, say $N$, Gaussian functions of identical width in either
direction, with first order $\tau$-dependent forward cumulant,
$\mu_i$, and weights, $c_i$, regulated by a generalization of the
Crooks theorem-based Eq. \ref{eq:jarzy2}:
\begin{eqnarray}
\Delta G_{1 \rightarrow 0} & = & -k_{B} T \ln \sum_i^N c_ie^{-\beta(\mu_i -\frac{\beta\sigma_i^2}{2})} 
\label{eq:gauss2} 
\end{eqnarray}
In the above equation, the forward $c_i$ weights satisfy the
constraint $\sum_i c_i=1$ and the reverse first order cumulants,
$\nu_i$ and weights, $d_i$, are related to the forward counterpart by
\begin{eqnarray}
\nu_i & = & \mu_i -\beta\sigma_i^2 \label{eq:nui}\\
d_i & = & e^{\beta\Delta G} e^{-\beta (\mu_i -\frac{\beta\sigma_i^2}{2})}  c_i 
\label{eq:di}\
\end{eqnarray}
In other words, the Crooks theorem, Eq. \ref{eq:ct}, imposes that, if
in one direction of the $\tau$-NE process the work distribution
happens to be given by a combination of $N$ normal distributions,
somehow connected to the existence of a manifold of free energy
basins, it must be so in the reverse process as well, albeit with
different combination coefficients given by Eq. \ref{eq:di}.
Eqs. \ref{eq:gauss2}-\ref{eq:di}, with $N=2$ explains surprisingly
well the striking asymmetry observed in systems where one direction of
the $\tau$-NE experiment (forward and/or reverse) envisages the
entrance in a funnel, like in the folding of a small
poli-peptide\cite{procacci06,Procacci10,Procacci2015} or, possibly, in
the docking of a drug on a receptor.  To see why in this latter case,
suppose that on one end of the $\tau$-NE process we have {\it only
  one} possible free energy basin (say the uncoupled state R +
(L)$_{\rm gas-phase}$ at $\lambda=0$), and on the other end (say the
coupled state RL at $\lambda=1$) {\it one} of the many basins has a
disproportionate Boltzmann weight with respect to weight of the others
all lying several $k_BT$. According to Eq. \ref{eq:onepose2}, the
overall weight of these secondary poses is given by
$W_s=\sum_{i=2}^{N_p} e^{-\beta \Delta G_{1}^{i}}$.  Then, provided
that $1/N_{\tau}> W_s$, {\it all} $N_{\tau}$ trajectories, starting
form the equilibrium fully coupled state with $\lambda=1$, should
include sub-states sampled only in the principal basin.  All these
$\tau$-NE trajectories, starting form the principal pose, end up into
the same state corresponding to the single free energy basin at
$\lambda=0$ of the free receptor and of the unbound ligand.  The
resulting forward work distribution should hence appear {\it quasi}
Gaussian with a $\tau$-dependent dissipation $\beta \sigma^2(\tau)/2$
and with inappreciable contamination on the left tail
of the distribution due to normal components related to the so-called
``shadow states''.\cite{Procacci2015} These shadows states can be only
explored and perceived as end $\tau$-NE states in the {\it reverse}
process where, \textcolor{Red}{for short $\tau$, most} of the final
$\tau$-NE poses would be clearly sub-optimal. As stated in
Ref. \cite{Procacci2015}, because of the mathematical structure of the
Crooks theorem for Gaussian mixtures, shadow states in the $\tau$-NE
reverse process undergo exponential amplification (see
Eq. \ref{eq:di}). From a physical standpoint, in the reverse process,
starting from the single-basin state, the components of the arrival
multi-basins thermodynamic state can be explored and detected because
of the extra energy provided by the dissipation that allows to
overcome the barriers between the basins.
\subsection{Standard state correction (SSC) in non equilibrium
unrestrained alchemical simulations} 

As previously discussed, unconstrained {\it reversible} DAM provides a
dissociation free energy $\Delta G_{\rm sim}=\Delta G_{\rm RL} -
\Delta G_{\rm L}$ that in principle should depends on the box volume
via Eq. \ref{eq:ssc}, but in the practice results in many cases
apparently independent on
it.\cite{Jorgensen85,pande06,Yamashita09,Yamashita2015} It has been
argued\cite{gilson97,karplus03,Roux09} that such apparent independence
on the simulation conditions arises since it is difficult to reach
full convergence in a simulation time of the order of the nanosecond
at small $\lambda$'s where the ligand may leave the binding site and
start to explore orientationally and translationally disordered
unbound states. In effect, the two decoupling processes, leading to
$\Delta G_{\rm RL}$ and $\Delta G_{\rm L}$, are both performed in the
same way: one must switch off the ligand interactions from
environments of comparable atomic density and having a common maximum
distance range of the order of 10:15 \AA. This given, it seems quite
unreasonable that in just one of these processes, the annihilating
free energy is so dependent on the volume or on the time scale of the
simulation.  The stubborn apparent independence of the computed
decoupling free energy for the bound state $\Delta G_{\rm RL}$ on the
volume of the simulation box and on the length of the simulations has
often lead\cite{Yamashita09,Yamashita2015,pande06} to essentially
identify $\Delta G_{\rm DAM}=\Delta G_{\rm RL} - \Delta G_{\rm L}$
with $\Delta G_{0}$ itself, even negating the very
existence\cite{Yamashita09} of the standard state correction
Eq. \ref{eq:ssc}. The mystery in DAM unrestrained simulations
involving the inability of detecting a measurable dependence of
$\Delta G_{\rm DAM}$ on either $V_{\rm box}$ or the simulation time,
eventually leaded to the development of the DDM reversible
theory,\cite{gilson97} where, in the annihilation of the ligand in the
complex, $V_{\rm ref}$ and $\xi_{\rm ref}$ are imposed using a biasing
potential impacting on the standard state correction via
Eq. \ref{eq:ssc}.  Tight biasing potentials allow a safe sampling at
any $\lambda$ in most cases within nanoseconds of simulation at the
price of artificially modifying the receptor exclusion zone, possibly
inhibiting the access to important part of $V_{\rm site}$ contributing
significantly to the integral of Eq \ref{eq:vsite}.  For infinitely
loose biasing potential, DDM clearly must coincide with DAM, provided
that we set $V_{\rm ref}=V_{\rm box}$ and $\xi_{\rm
  ref}=8\pi^2$,\cite{Roux09} leading to the relation $\Delta G_{\rm
  DAM}=\Delta G_{0} - k_BT \ln \left ( \frac{V_{\rm ref}} {V_{0}}
\right )$, in principle correct, but consistently contradicted in the
simulation practice.  As we shall see in the following, the
identification of $\Delta G_{\rm DAM}$ with a volume independent
system quantity related to $\Delta G_{0}$ rather than $\Delta G_{\rm
  sim}$ in DAM can be assumed to be legitimate if unrestrained DAM is
conceived as a {\it non equilibrium} experiment (NE-DAM), with many
long NE trajectories producing a very narrow, apparently Gaussian,
work distribution, with a {\it shadow component} at a much lower
energy dependent on the simulation volume, obeying the Crooks
theorem-derived Eq. \ref{eq:gauss2}. In other words, the true
volume-dependent value of $\Delta G_{\rm sim}$ in DAM is {\it
  computationally unattainable} in a {\it single simulation}.
   
In principle, one could straightforwardly implement a NE variant of
DDM using a restrained potential that keeps the volume in the binding
site during the decoupling process, as it occurs in reversible DDM.
However, a restraint potential in NE-DAM is not necessary neither
desirable.  As previously stated, the restraint potential is
introduced in equilibrium alchemical transformation to limit the
sampling of the ligand accessible ${\bf r},\Omega$ space, in order to
make the transformation reversible. In NE-DAM or FS-DAM the decoupled
states in the semi-open interval (1,0] are by definition non equilibrium states
with no specific requirements of sampling, except for those dictated
by the initial bound configurations at $\lambda=1$ (the only states
sampled at equilibrium) and by the time $\tau$ of the NE
experiments. Moreover, in the annihilation of the ligand in the bound
state, the final available translational and rotational volumes for
the ligand depend on the time $\tau$ of the NE
simulations.\cite{fsdam} Borrowing the notation from the equilibrium
relation Eq. \ref{eq:ssc}, we define these NE volumes as $V(\tau)$ and
$\xi(\tau)$.  Given a forward $\tau$-NE transformation,
Eq. \ref{eq:ct} applies if the process can be inverted. While for the
ligand in the bulk the decoupling process can be straightforwardly 
inverted with a $\tau$-lasting inverted-schedule growth process, for
the ligand in the complex, the reverse (growth) process is more
elusive.  As stated in Ref. \cite{fsdam}, an hypothetical reverse
process of the same duration $\tau$ with inverted time schedule from
the decoupled state of the complex to the fully coupled state should
be performed by switching on first the dispersive-repulsive
(soft-core) potential of the ligand and then the electrostatic
interaction, with the gas-phase decoupled ligand in initial positions
and orientations relative to the receptor sampled randomly from the NE
volumes $V(\tau)$ and $\xi(\tau)$ found in the forward
transformation. By virtue of the Crooks theorem. Eq. \ref{eq:ct}, this
reverse work distribution $P(-W_{0\rightarrow 1})$ must cross the
forward counterpart at a {\it $\tau$-dependent} free energy $\Delta
G_{\rm RL}(\tau)=\Delta G_{\rm RL}(V(\tau),\xi(\tau))$.  To get rid of the
$\tau$-dependence in $\Delta G_{\rm RL}$ we can imagine to do the forward NE
transformation in two step. In the first stage, we switch off the
ligand-environment interactions almost completely up to an arbitrarily
small $\lambda_{\tau}=\delta \lambda$ in the time $\tau$,  obtaining
basically the same forward work distribution $P(W_{1\rightarrow 0})$
of the complete $\tau$-NE process.  Thanks to the soft-core potential,
when $\lambda$ is infinitesimally small, the ligand does not sense
anymore the environment and starts to move ballistically in a random
direction with random translational/rotational velocities.  In a
second step, doing practically no work, we finally switch off the residual
interaction in a time $\tau_{\rm box}$ long enough so that the
$N_{\tau + \tau_{\rm box}}$ end states get randomly distributed in the
whole simulation box.  The reverse ${\overline{\tau_{\rm
      box}+\tau}}$-NE process, in this case, is essentially equivalent
to the switching on of the ligand in a time $\tau$ starting from a
random position and orientation within the simulation box. With this
time protocol, $\Delta G_{\rm sim}$, like in DAM, must be a function
only of $V_{\rm box}$, no longer depending on $\tau$, so that
\begin{equation}
\Delta G_{\rm sim} = \Delta G_{\rm RL} - \Delta G_{\rm L} = \Delta G_{0} - k_B T \ln \frac{V_{\rm box}}{V_0}
\label{eq:nessc}
\end{equation}  
In the reverse ${\bar{\tau}}$-NE process, in most of the
NE-trajectories, the ligand is switched on in the bulk solvent or in a
sub-optimal random pose on the receptor surface yielding a mean work
(with inverted sign) that is substantially smaller than the mean work
obtained in the forward transformation. The distance between the
forward and reverse distribution is of the order of the $V_{\rm box}$
dependent dissociation free energy so that, for tight binding ligand,
$P(W_{1\rightarrow 0})$ and $P(-W_{0\rightarrow 1})$ have a negligible
overlap.\cite{fsdam} 
\begin{figure}[h!]
\includegraphics[scale=0.35]{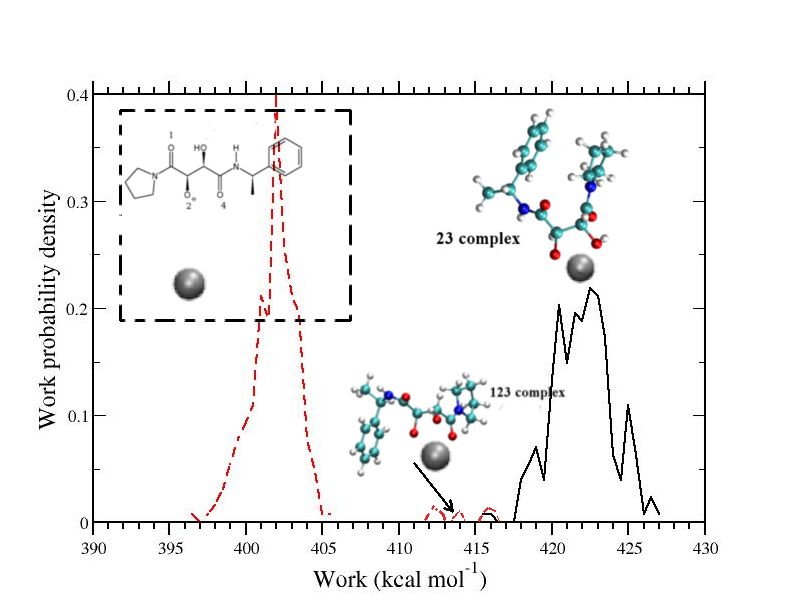}
\caption{Forward (solid, black) and reverse (dashed, red) work distribution in the
  Zn(MBET306)$^{-1}$ complex in water calculated using fast switching NE simulations.}
\label{fig:mbet}
\end{figure}
In the Figure \ref{fig:mbet}, I report, as an illustrative example,
the forward and reverse work distribution in a real unrestrained fast
switching NE-simulations, that is the annihilation/growth of the
Zinc(II) cation in the Zinc(II)-MBET306$^{+}$ complex in explicit
water in standard condition in a cubic MD box of volume $V_{\rm
  box}\simeq 15000~\AA^3$. The distribution were obtained using 256
annihilation/growth runs each lasting 90 ps. Further details on the
simulations are given in Ref.\cite{fsdam} On the right (solid black
line), we have the annihilation work distribution $P(W_{1 \rightarrow
  0})$ of the principal bound state of the Zn-MBET306$^{+}$ bound
species. On the left (dashed red line), I report the reverse
distribution $P(-W_{0 \rightarrow 1})$ corresponding to the growth of
the Zinc(II) cation from a random position in the MD box in presence
of the MBET306$^{-1}$ receptor, where the most likely final NE-state
corresponds to an unbound Zinc(II) in the bulk solvent.  The small
features of $P(-W_{0 \rightarrow 1})$, at about 410:415 kcal
mol$^{-1}$ with overall weight proportional to with $V_{\rm
  site}/V_{\rm box}$, corresponds to a few trajectories yielding a
work that is related to the secondary poses of the Zn(II) on the
MBET306$^{-}$ anion.

The pattern shown in Figure \ref{fig:mbet} closely resembles that
seen in systems where one direction of the $\tau$-NE experiment
envisages the entrance in a funnel, like in the folding of a small
poli-peptide.\cite{Procacci2015} The entrance in the exclusion zone
(that for the case of the Zn(MBET306)$^{+}$ complex reported in Figure
\ref{fig:mbet} corresponds roughly to the volume surrounding the
central tartrate core of the molecule), via fast-growth from randomly sampled positions
in a volume that is much larger than $V_{\rm
  site}$,  is a far more dissipative process than the alchemically
driven escape of the ligand from the binding site.  Based on this
analogy, we make the Ansatz that the forward decoupling work
distribution $P(W_{1\rightarrow 0})$ \textcolor{Red}{for tightly bound
ligand receptor system} is made of essentially of one
principal normal distribution relative to the starting stable pose of
the ligand in the exclusion zone, $N(W,\mu,\sigma)$, and by a negligibly
small volume-related work distribution $N(W,\mu_{\rm box},\sigma_{\rm
  box})$ due to sub-optimal poses or unbound states that could be
detected in the reverse recoupling process:
\begin{eqnarray}
P(W_{1\rightarrow 0}) & = & (1-c)N(W,\mu,\sigma) + cN(W,\mu_{\rm
  box},\sigma_{\rm box}) \label{eq:ness0} 
\end{eqnarray}
In Eq. \ref{eq:ness0} we have therefore that $c\ll 1$. With this
regard, it is important to realize that, the second normal component
depending on $\mu_{\rm box}$, while negligible in shaping the forward
distribution, because of the Crooks theorem Eq. \ref{eq:ct} {\it gets
  exponentially amplified in the reverse process} (see
Eq. \ref{eq:di}). We further assume, in Eq.  \ref{eq:ness0}, that $0 <
\mu_{\rm box} < \mu$ and $\sigma\simeq \sigma_{\rm box}$. The first
assumption simply says that the non covalent complex exists, and hence
one must do work to switch off the interaction with the environment
and that this work must be larger than the mean work $\mu_{\rm box}$
done to switch off the interaction with the environment when the
ligand may no longer be in the exclusion zone. \textcolor{Red}{The
  second simplifying assumption implies no loss of generality\cite{Procacci2015} and 
is based} on the reasonable expectation that the mean dissipation,
$\beta\sigma^2/2$, depends in essence on the particle density in the
given thermodynamic conditions and that therefore $\sigma$ should be weakly
dependent on the environment surrounding the ligand.  Given the
forward distribution Eq. \ref{eq:ness0}, the Crooks theorem,
Eq. \ref{eq:gauss2}, imposes that the reverse distribution,
\begin{eqnarray}
P(-W_{0\rightarrow 1}) & = & d~ N(W,\nu,\sigma) +
(1-d)N(W,\nu_{\rm box},\sigma_{\rm box}) 
\label{eq:ness}
\end{eqnarray}
\begin{figure}[h!]
  \caption{Forward/decoupling (left) and reverse/re-coupling (right)
    NE alchemical process in drugs receptor system. Six possible
    outcomes of NE trajectories are shown.  The receptor is depicted
    in blue with the a square-shaped exclusion zone to allocate the
    ligand corresponding to a single box unit of the 2D-grid.  The drug is
    red when is fully interacting with the environment and is light
    red when is in the decoupled state. In the forward process on the
    left, all the $N_{\tau}$ equilibrium starting configurations are
    in the bound state and the final states the ligand ends up in a
    random position in the MD box.  In the reverse process, the
    decoupled ligand is initially randomly distributed in the box and
    ends in different unbound states or sub-optimal poses on the
    receptor.  The probability to end up in the correct bound state is
    given by the volume of the exclusion zone divided by the total
    volume of the MD box, i.e. to $1/N_{\rm grid}$.}
  \includegraphics[scale=0.24]{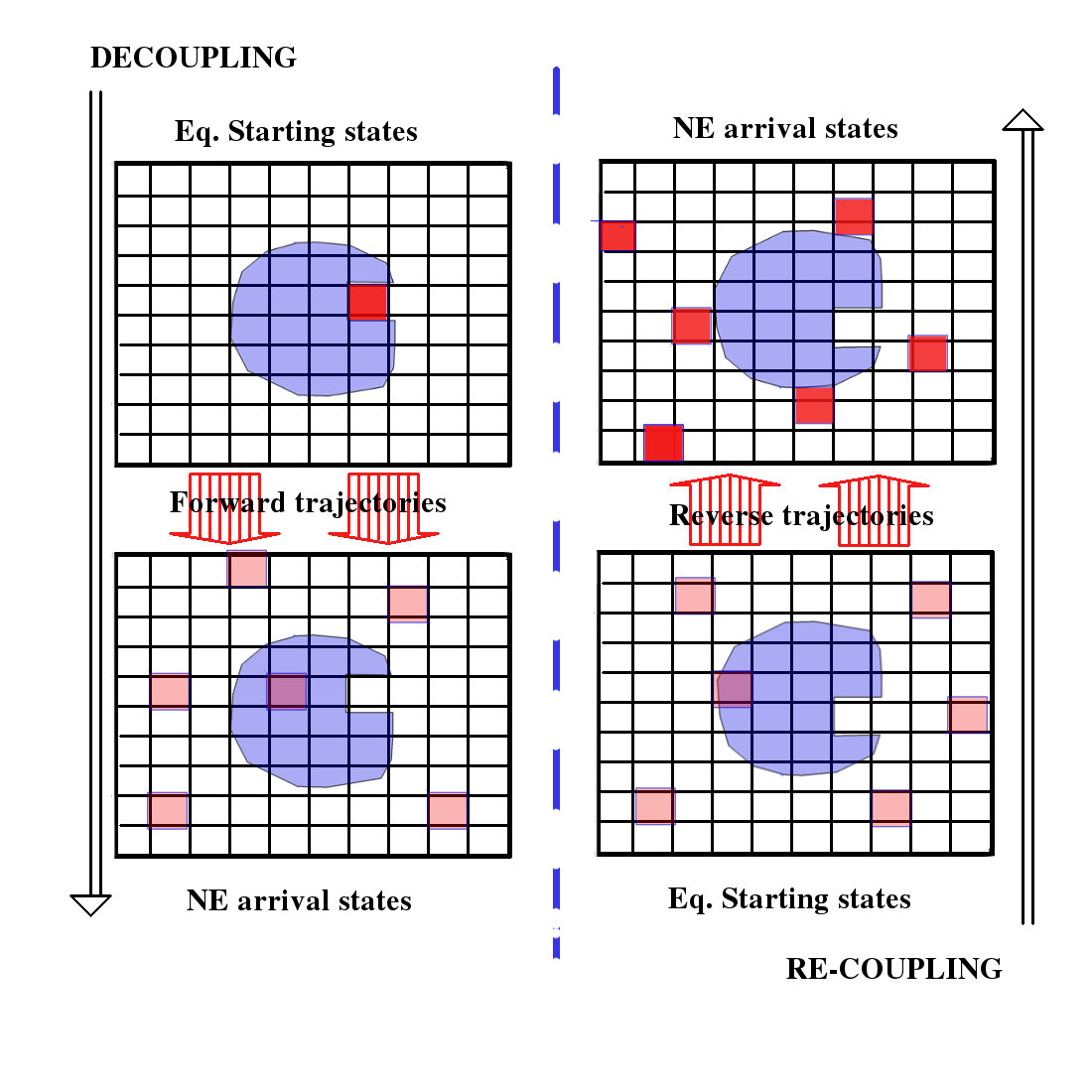}
\label{fig:ssc}
\end{figure}
is such that $\nu=\mu - \beta \sigma^2$ and $\nu_{\rm
  box}=\mu_{\rm box} - \beta \sigma^2$ (See Eq. \ref{eq:nui}). The
weight $d$ of the reverse Gaussian normal component with mean $\nu$ in
Eq. \ref{eq:ness} equals the probability of growing the ligand in the
exclusion zone form a random position in the volume $V_{\rm box}$, i.e
\begin{equation}
d= V_{\rm site}/V_{\rm box}.
\label{eq:dcoeff}
\end{equation} 
If $V_{\rm box} \gg V_{\rm site}$, as it occurs in the simulation
practice, then $d$ is small and the principal component of the reverse
process is $N(W,\nu_{\rm box},\sigma_{\rm box})$.
The volume dependent free energy $\Delta G_{\rm RL}$ is found at the
crossing point of the $\mu$-related forward and reverse Gaussian
component:
\begin{eqnarray}
\Delta G_{\rm RL} & = & \mu - \frac{1}{2} \beta \sigma^2 + k_BT \ln \frac{d}{1-c}
\nonumber \\
                  & \simeq & \Delta G_{x} + k_BT \ln d  \label{eq:dgx}
\end{eqnarray}   
where in the last equation we have exploited the fact that $c\ll1$ and
we have defined $\Delta G_{x} = \mu - \frac{1}{2} \beta \sigma^2$.
For Gaussian NE processes, the quantity $\mu - \frac{1}{2} \beta
\sigma^2$ should be invariant with respect the duration time $\tau$ of the
experiment, always yielding the minimum reversible work to do the
transformation. As a matter of fact, if we make $V_{\rm box}$ larger, we need to set
$\tau_{\rm box}$ larger but we clearly have no impact on the mean work $\mu$
done up to $\tau$. So $\Delta G_{x}$, unlike the crossing point
$\Delta G_{\rm sim}$, {\it does not depend on the box volume}.
Using Eq. \ref{eq:dcoeff} we finally find 
\begin{eqnarray}
\Delta G_{\rm RL} & = & \Delta G_{x} +   k_BT \ln \frac{V_{\rm site}}{V_{\rm box}} \\
                  & = & \Delta G_{x}  +  k_BT \ln
\frac{V_{site}}{V_0}  - k_B T \ln \frac{V_{\rm box}}{V_0}
\label{eq:dgsim2}
\end{eqnarray}  
By subtracting on both side of Eq. \ref{eq:dgsim2} the volume
independent solvation free energy of the ligand $\Delta G_{\rm L}$ and by
using Eq. \ref{eq:nessc}, we finally find that the standard
dissociation free energy in NE alchemical transformation is given by
\begin{equation}
\Delta G_0 = \Delta G_{x} - \Delta G_{\rm L} +  k_BT \ln
\frac{V_{site}}{V_0}
\label{eq:dam}
\end{equation}
In deriving Eq. \ref{eq:dam} from NE theory, DAM theory is somehow
vindicated. The annihilation free energy of the complex in DAM may be
thought as being derived from a high number of slow (ns time scale)
{\it quasi-equilibrium} trajectories yielding a very sharp and
normally distributed $\Delta G_{x}$ plus a {\it shadow state} that
could be visible only if one does the reverse reaction, i.e. the
switching on of the ligand in a random position of the MD box, {\it in
  presence of the receptor}. In the context of NE thermodynamics, the
DAM free energy $\Delta G_{x}$ is indeed a system dependent quantity
as conjectured in Ref. \cite{Yamashita09,pande06} that needs only to
be shifted to match the SSC reference value by the $k_BT \ln
\frac{V_{site}}{V_0}$. This correction for drug size ligand, is
actually very small. Using value of 10 \AA$^3$ as the mean volume per
atom in condensed phases in standard conditions, we may estimate
$V_{\rm site}$ using the volume of the ligand itself, obtaining a SSC
correction ranging from -0.7:0.1 kcal mol$^{-1}$.\cite{pande06}

\section{Competitive poses and conformational sampling in non equilibrium simulations} 

We have seen that FS-DAM and DAM can be both embedded in the context
of non equilibrium transformations. FS-DAM and DAM differ only in the
speed of the NE process, fast in FS-DAM, very slow in DAM.  In both
cases the distribution $P(W_{1 \rightarrow 0})$ for the annihilation
of the ligand is normal with insignificant contamination by normal
components of shadow states due to poses {\it outside the exclusion
  zone} or to {\it unbound states}. The distributions relative to
these shadow states are exponentially amplified via Eq. \ref{eq:di} in
a hypothetical (and unnecessary) reverse process where we grow the gas-phase ligand in
a random position in the MD box and with random orientation with
respect to the receptor, producing, when dealing with tight-binding
ligand, a main normal component $N(W,\nu_{\rm box})$ with no overlap
with the forward Gaussian component $N(W,\mu)$ as shown in the example
reported in Figure \ref{fig:mbet}.
\begin{figure}[h!]
  \caption{Schematic representation of the forward and reverse work
    distribution in the alchemical decoupling process for a ligand
    with one principal pose and with a secondary orientational pose
    (see text for details). Normal components related to unfavorable
    ligand-receptor free energy basins are exponentially amplified
    in a hypothetical reverse process (dashed, red line).  Assuming that
    the volume of the exclusion zone is such that $V_{\rm site} \simeq
    V_{0}$, the example show a possible $P(W_{1 \rightarrow
      0}),~P(-W_{0 \rightarrow 1})$ diagram for a ligand with a
    dissociation energy in the order of 20 $k_BT$.}
  \includegraphics[scale=0.33]{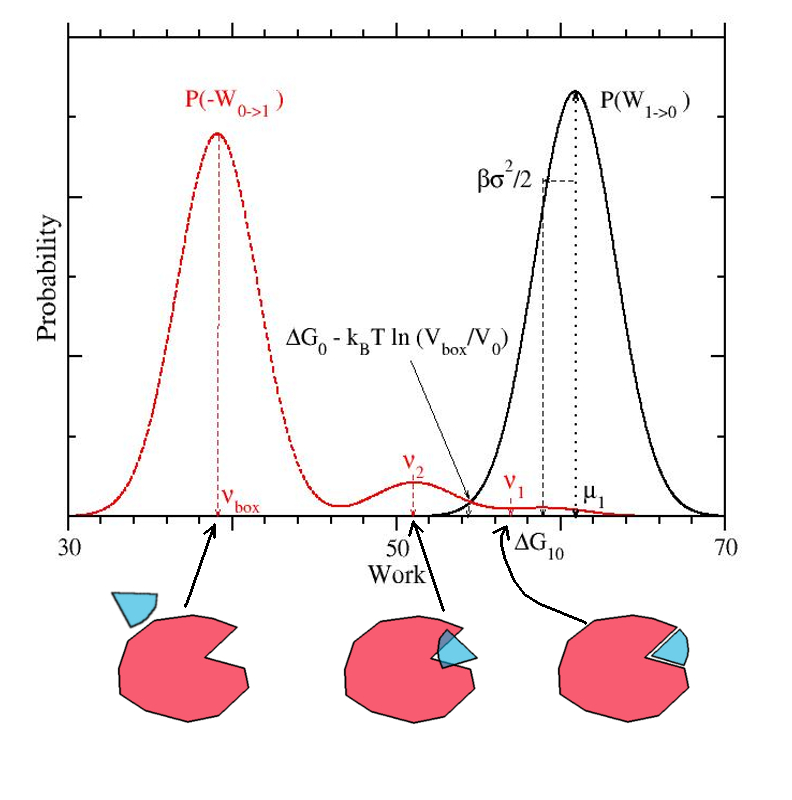}
\label{fig:Fig2}
\end{figure}
We have also seen, in the preceding section that, given a spread of
the work distribution of few kcal mol $^{-1}$ for speeds of the
decoupling process lasting in the order of 50:300 picoseconds, few
hundreds of $\tau$-NE trajectories are sufficient for getting an
accuracy within 0.5 kcal mol$^{-1}$ in the dissociation free
energy.\cite{fsdam} In the scheme reported in Figure \ref{fig:ssc} we
have implicitly assumed that the bound state free energy is
independent of the orientation of the ligand in the binding site. In
reality, the ligand could be found in the exclusion zone with, e.g.,
several competing and mutually exclusive orientational poses (or free
energy basins), with one of such poses being much more favorable than
all the others (see Eq. \ref{eq:onepose2}).  A minimum relative free
energy difference such \textcolor{Red}{that ${\rm Min}_{i \ne 1}(\Delta
G^{i}_{1})>3.72$} kcal mol$^{-1}$ translates in a probability ratio
$P_{i}/P_{1} < 1/512$ and is hence sufficient to exclude {\it all} the
conformations due to the secondary poses $i=2..N_{p}$ from the pool of
the few hundreds starting states of the bound complex randomly sampled
out an equilibrium distribution. It follows that the apparent
distribution due to the $N_{\tau}$ trajectories is again, in essence,
normal, although is now made (in the limit $N_{\tau} \rightarrow
\infty$ or for averages over infinite \textcolor{Red}{non overlapping} bootstrap $N_{\tau}$ samples) of
      {\it three components}, namely that due to the principal pose,
      that due to the secondary poses in the exclusion zone with
      weight $c_2=e^{-\beta \Delta G^{2}_{1}}< \frac{1}{512}$ and the
      shadow state due to the sub-optimal poses on the receptor
      surface outside the exclusion zone or in the solvent with even
      smaller weight $c_{\rm box} \simeq e^{-\beta \Delta G_0}$:
\begin{eqnarray}
P(W_{1\rightarrow 0}) & = & (1-c_2-c_{\rm box})N(W,\mu_1,\sigma) +
c_2N(W,\mu_{2},\sigma)+ \nonumber \\ 
& + &  c_{\rm box} N(W,\mu_{\rm box},\sigma) \label{eq:ness3}. 
\end{eqnarray}
where, for the sake of simplicity, we have assumed a pose independent
spread/dissipation $\sigma$ for all $\tau$ lasting NE
annihilation/growth processes.  As already discussed, the Gaussian
nature of the annihilation work distribution is somehow guaranteed by
the speed (few tens of few hundreds of picoseconds) with which the
alchemical decoupling is carried on allowing only marginal mixing
between the underlying free energy basins at the intermediate NE
$\lambda$ states.  This is clearly at variance with reversible
transformations, especially when implemented with $\lambda$-hopping
schemes, that are introduced precisely to favor the canonical mixing
all along the alchemical coordinate. It should also be noted that
$\lambda$-hopping schemes, based on probabilistic criteria for the
$\lambda$ dynamics, are either at convergence or
they are incompatible with NE theory since they make the annihilation
process not invertible.  We have seen that the $\nu_{\rm box}$ related
coefficient, $1-d$, in the reverse distribution of Eq. \ref{eq:ness}
gets exponentially amplified via the Crooks theorem-derived
Eq. \ref{eq:di}. By the same token, in a hypothetical reverse process,
the normal components due to secondary poses in the exclusion zone are
exponentially amplified via Eq. \ref{eq:di} so that the bound-stated
related minor peak at $\nu$ integrating to $V_{\rm site}/V_{\rm box}$
(see Eq. \ref{eq:ness} and Eq. \ref{eq:dcoeff}) {\it gets split} in
a left-most peak due to the manifold of secondary poses and to a
smaller peak due to the principal pose whose height is proportional to
the ratio $\xi(\Omega)/8\pi^2$ where $\xi(\Omega)$ is the fraction the
domain \textcolor{Red}{$\{{\bf r}, \Omega \}: I({\bf r}, \Omega ) = 1$.} On the
overall, the Crooks theorem-related reverse distribution is of the
form
\begin{eqnarray}
  P(-W_{0\rightarrow 1}) & = & d_1~N(W,\nu_{1},\sigma) +
  d_2N(W,\nu_{2},\sigma)+ \nonumber \\ 
& + & (1-d_1-d_2)N(W,\nu_{\rm box},\sigma) \label{eq:rness3}. 
\end{eqnarray}
with $d_1<d_2<(1-d_1-d_2)$.  In Figure \ref{fig:Fig2} these concepts
are schematized. The reverse distribution (dashed, red line) exhibits
a principal left-most peak, $\nu_{\rm box}$, due to the sub-optimal
poses off the binding site, an intermediate peak $\nu_{2}$ due a
wrongly oriented poses in the binding site and a weak component due to
the primary pose $\nu_{1}=\nu$ that is strongly overlapping with the
forward apparently single component annihilation
distribution. Assuming for the sake of simplicity and without loss of
generality that $V_{\rm site} \simeq V_{0}$ such that $\Delta G_{x}
\simeq \Delta G_{0}$, the crossing point of the forward and reverse
distribution is again, as in Eq. \ref{eq:dgx}, at the point $\Delta
G_{0} + k_BT \ln d$ and again the free energy $\Delta G_{0}$ can be
computed using the single Gaussian unbiased estimate of
Eq. \ref{eq:jarzy2}. In order to convey the concept, the weight of the
components due to the bound states, $\nu_1$ and $\nu_{2}$ in the 3-G
reverse distribution of Eq. \ref{eq:rness3} have been greatly
exaggerated. Actually, the ratio $V_{\rm site}/V_{\rm box}$, i.e. the
overall weight of the bound states in the unrestrained reverse
distribution for real drug-receptor system is expected in the range
$V_{\rm site}/V_{\rm box}$=0.01:0.001, implying that only few
trajectories out of hundreds could produce a work corresponding to a
bound state, exactly as observed in the growth of the Zinc(II) cation
in presence of the MBET306$^{-}$ anion (see Figure \ref{fig:mbet}).
Besides, the peak relative to the principal pose, $\nu_1$, is
exponentially abated via Eq. \ref{eq:di} so that basically none out of
few hundreds reverse NE trajectories is expected to yield a
\textcolor{Red}{work (with inverted sign) 
falling near} the forward distribution $P(W_{1\rightarrow 0})$.  Based
on the reverse process shown in Figure \ref{fig:mbet} for a simple
atomic ligand, we conclude that a hypothetical reverse process in
unrestrained NE DAM in real drug-receptor systems would systematically
produce a forward and reverse distributions separated by a large gap,
related to the dissociation energy itself, making bidirectional
estimates such as Bennett acceptance ratio
unreliable.\cite{Procacci2015} The principal-pose assumption in the
bound complex, leading to Eq. \ref{eq:onepose2}, constitutes the
thermodynamic basis for molecular recognition.  Most importantly, the
existence of a pose with overwhelming Boltzmann weight in the complex
implies a nearly Gaussian distribution in the $\tau$-NE decoupling of
the ligand, allowing a reliable and unbiased estimate of the
annihilation free energy $\Delta G_{\rm RL}$ to be obtained via the
simple, unbiased Gaussian estimate Eq. \ref{eq:jarzy2}. Such principal
pose must of course be known from the start to be able to sample the
equilibrium initial states at $\lambda=1$ in the corresponding free
energy basin via standard molecular dynamics. Secondary poses can be
checked in a similar manner, in a separate and independent NE
experiment by using initial states {\it all} sampled in the
corresponding secondary free energy basins. Again, if the NE
simulations are so fast that only marginal mixing occurs among poses
during the decoupling of the ligand, then the absolute dissociation
free energy, $\Delta G_{i0}$ of the $i$-th secondary pose can also be
determined using a simple Gaussian estimate, yielding as a trivial
byproduct, the relative free energy difference $\Delta G_1^{i} =
\Delta G_{i0}-\Delta G_{10}$, i.e the Boltzmann weight of the $i$-th
pose relative to the principal pose, $\exp(-\beta(\Delta G_{i0}-
\Delta G_{10}$). Such an approach has been used successfully in
Ref. \cite{fsdam} to derive the overall binding constant in water of
the complex of the Zinc(II) with the MBET306$^{-1}$ anion, an inhibitor
of the Tumor necrosis factor $\alpha$ converting enzyme. In that study,
Sandberg {\it et al.}  examined via NE unrestrained unidirectional
simulations more than ten different poses of the cation on the
tartaric moiety of MBET306$^{-}$.
 
\section{Fast Switching calculation of solvation energies $\Delta G_{\rm L}$ }
  
In the complex, the preparation of the equilibrium starting states is
an easy one. The ligand, by filling the exclusion zone of the
receptor, inhibits its conformational motion and that of the protein
residues delimiting the binding site, hence reducing the
conformational entropy of the complex. 
\begin{figure}[h!]
  \caption{Fast switching forward (dashed black) and reverse (dashed red) work distributions
    obtained for N-Elte378 in water ($N_{\tau}=512$,
    $\tau$=270 ps. The free energy was evaluated using Eq. \ref{eq:gauss2} assuming two normal components. Error bars reported on the fitted distributions (solid
    lines) were computed by bootstrapping samples of 256 works. 
}
  \includegraphics[scale=0.35]{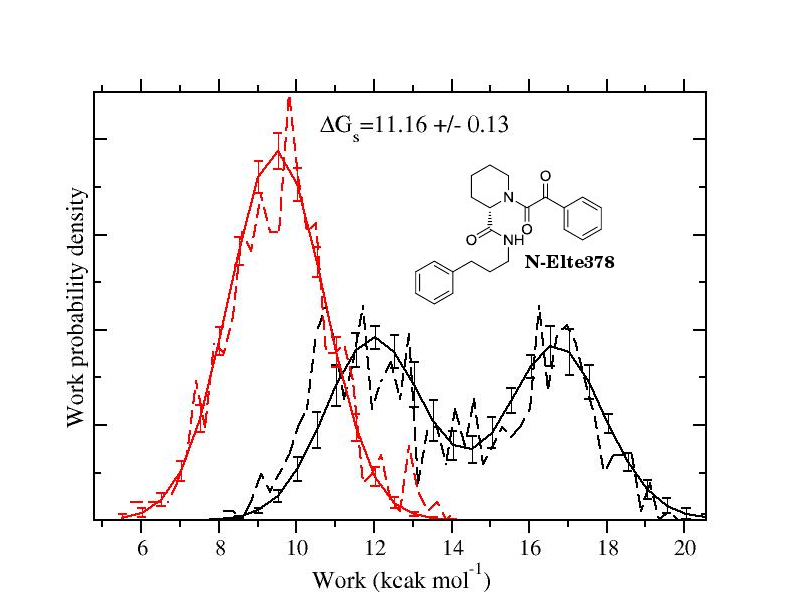}
\label{fig:elt}
\end{figure}
Once the $N_{\tau}$ NE independent trajectories from these states are
launched, unlike in $\lambda$-hopping reversible DDM, we are no longer
concerned with equilibrium sampling. Each $\lambda$-driven trajectory
ends up irreversibly in the NE final decoupled state producing a mean
work that depends chiefly on the enthalpy of the starting equilibrium
state and not on the intermediate states that are rapidly crossed.
FS-DAM, like DAM, needs also to annihilate the ligand in the bulk to
get $\Delta G_{\rm L}$ and hence $\Delta G_{0}$ via Eq. \ref{eq:dam}.
While the calculation of $\Delta G_{\rm L}$ is computationally far
less demanding than the decoupling free energy of the bound state, the
starting equilibrium states of the free ligand in bulk, especially
when the ligand exhibits competing conformations of comparable free
energies, should be prepared with the due care. I report as an
illustrative example the case of the N-Elte378
[(2S)-1-(2-oxo-2-phenylacetyl)-N-(3-phenylpropyl)
  piperidine-2-carboxamide], a tight binding synthetic ligand of the
immunophilin FKBP12.\cite{bizzarri13}. N-Elte378, a conformationally
disordered molecule, can be characterized in water by a competition
between extended and compact conformations (see Figure 7 of
ref. \cite{bizzarri13}), the latter being stabilized by persistent
stacking interaction involving the two terminal phenyl moieties. The
starting equilibrium configurations of N-Elte378 in water for the fast
switching calculation of $\Delta G_{\rm L}$ in bulk are taken from a
Hamiltonian Replica Exchange simulation with torsional tempering
reported in Ref. \cite{bizzarri13}.  Simulations details and methods
can be found in Ref. \cite{bizzarri13}.  The fast annihilation
(forward) works were obtained running, in a single parallel run, 512
NE-trajectories lasting 270 ps. During the NE process, the solute is
linearly discharged in the first 120 ps, followed by the switching off
of 2/3 of the dispersive-repulsive interactions up to 150 ps. In the
last 120 ps the residual Lennard-Jones interaction is finally switched
off, using a soft-core regularization to avoid numerical instabilities
near $\lambda=0$.\cite{Beutler99} The fast growth (reverse) work from
gas-phase N-Elte378 were collected with inverted time schedule using
again 512 NE trajectories. The parallel computations were done using
the fast switching alchemy version of the ORAC
code\cite{Procacci14,Marsili10} in less than one wall-clock time hour.
In Figure \ref{fig:elt},  I report the computed forward and reverse
work distributions for N-Elte378 in water (dashed lines) along with
the fitted distributions using Eq. \ref{eq:gauss2} with two normal
components (solid lines). Due to the complex conformational manifold,
and because of the significant mixing between conformations during the
270 ps decoupling process, the annihilation work distribution in
solvated N-Elte378 does not appear as a simple normal distribution,
roughly reflecting the bi-modal structure observed in the probability
distribution of the distance between the two terminal phenyl moieties
(see Figure 7 of Ref. \cite{bizzarri13}).  Still, Eq. \ref{eq:gauss2}
explains very well the observed strikingly asymmetrical forward and
reverse distributions, that were fitted assuming two normal components
($N=2$ in Eq. \ref{eq:gauss2}). The errors bars on the fitted
distributions and on the hydration free energy are computed by block
bootstrapping the collection of 512 work using 40 samples with 256
works. The bidirectional free energy computed using the Bennett
acceptance ratio using the forward and reverse 512 works is computed
at 11.02 $\pm$ 0.05 kcal mol$^{-1}$, comparing favorably with the
estimate of 11.16 $\pm$ 0.16 kcal mol$^{-1}$ based on
Eq. \ref{eq:gauss2}.

\section{Conclusions and perspective}

In this study I have revisited the theory of non covalent bonding in
the evaluation of the binding free energies in drug-receptor systems
from a non equilibrium perspective. I have shown that, in the context
of the alchemical approach, the dissociation free energy of the
complex can be effectively and accurately derived producing few
hundreds of non equilibrium unrestrained trajectories starting from
canonically sampled fully coupled bound states. The inherent Gaussian
nature of the probability of doing a work $W$ at the end of the fast
annihilation process allows to recover the decoupling free energy
using a very robust unidirectional unbiased estimate.  The fast
switching double annihilation estimate (FS-DAM) is based on the
assumption that the forward annihilation and the hypothetical reverse
growth work distributions of the ligand in the complex and in the bulk
are given by a mixture of normal distributions with weights regulated
by the Crooks theorem. The standard state correction, related to the
volume of the exclusion zone in the receptor, arises naturally in non
equilibrium alchemical transformations with no need for restraining
the motion of the ligand in the bound state. Non equilibrium
unrestricted alchemical transformations eliminate altogether the
necessity for canonical sampling at intermediate $\lambda$ states that
constitutes the major stumbling block in the reversible alchemical
approach.  In this regard, one of the most critical aspects in
reversible alchemical simulations, intimately related to the sampling
issue, is the need of minimizing the overall statistical uncertainty
of the free energy evaluation with respect to the alchemical protocol,
that is, of equalizing the contribution to the uncertainty across
every point along the alchemical path. In FS-DAM, equilibrium sampling
is required at one single point along the alchemical path, at the
fully coupled Hamiltonian. As a consequence, the accuracy of FS-DAM
free energies depends in a predictable way on the resolution of the
resulting work distribution, i.e. on the ratio of the spread of the
work distribution and on the number of NE independent trajectories.
The Crooks theorem-based estimate of the FS-DAM free energies relies
on the determination of the first two cumulant of a normal
distribution, whose variance is subject to the ancillary
$t$-statistics and is \textcolor{Red}{proportional to
$\sigma(\tau)/(N_{\tau})^{1/2}$ and $\sigma^2(\tau)/(N_{\tau})^{1/2}$,}
where $\sigma(\tau)$ is the $\tau$-dependent spread of the
distribution.  Reducing the number of NE trajectories by a factor of
$G$ amplifies \textcolor{Red}{the error on $\mu$ and $\sigma^2$} only by $G^{1/2}$
making FS-DAM estimates extremely robust and reliable even with a very
small number of sampling trajectories.\cite{Procacci2015}

As the NE-trajectories can be run independently, the FS-DAM approach
can be straightforwardly and efficiently implemented on massively
parallel platforms providing an effective tool for virtual screening
in the drug discovery process. In the applicative companion
paper\cite{paper2} of the present theoretical contribution, we apply
the FS-DAM technology to a challenging drug-receptor system, the
FKBP12 protein associated to the FK506 related ligands, comparing
performances and accuracy to the standard equilibrium approach.  In
that study \cite{paper2} we show that FS-DAM satisfactorily reproduces
the experimental dissociation free energies of several FK506-related
bulky ligands towards the native FKBP12 enzyme in a single massively
parallel run in matter of {\it few wall time clock hours} on a High
Performance Computing facility.  FS-DAM is finally used to predict the
dissociation constants for the same ligands towards the FKBP12 mutant
Ile56Asp. The effect of such mutation on the binding affinity of
FK506-related ligands is relevant for assessing the thermodynamic
forces regulating molecular recognition in FKBP12
inhibition. Moreover, the binding affinities of FK506-related ligands
for the Ile56Asp FKBP12 mutant are, to our knowledge, not yet available, exposing
our FS-DAM predictions to experimental verification. Anticipating the
results presented in Ref. \cite{paper2}, we summarize in Table \ref{tab:times}
performance and accuracy tests of the FS-DAM method compared to the
standard equilibrium approaches for the evaluation of the dissociation
constant of a drug-receptor pair in explicit solvent.  
\begin{table}[h!]
\begin{center}
\begin{tabular}{c|cccc}
\hline\hline
              &    $N_{\tau}$ & $N_{\lambda}$ &   Simulation time &  Mean error on                                     \\
              &                &               &  (ns per ligand)    &                $\Delta G_{0}$ (kcal mol$^{-1}$)    \\
\hline
FS-DAM  & 512   & n/a & 218   & 0.3 \\  
FS-DAM  & 256   & n/a & 149   & 0.7 \\  
FS-DAM  & 128   & n/a & 115    & 1.5 \\  
FEP\cite{Shirts_thesis}      &  n/a & 31 & 18000 & 1.5 \\
FEP\cite{Fujitani2005}      &  n/a & 33 & 400  & 4.5 \\
FEP/BAR\cite{Yamashita09}   & n/a  & 32 & 900  & 3.0 \\
FEP-restraint\cite{Wang06}  & n/a  & 25 & 250  & 1.5 \\
\hline\hline
\end{tabular}
\end{center}
\caption{Performances of NE FS-DAM and equilibrium FEP.
$N_{\tau}$ and $N_{\lambda}$ indicate the number of independent NE trajectories (applicable in FS-DAM only) and the number of $\lambda$ intermediate states 
(applicable in FEP only). All data refer to the FKBP12 receptor\cite{paper2} on per ligand basis. }   
\label{tab:times}
\end{table}
These results are fully detailed in Ref. \cite{paper2} and show that
FS-DAM outperforms FEP
approaches,\cite{Shirts_thesis,Fujitani2005,Yamashita09} both in terms
of precision/reliability and of CPU time.  The efficiency, simplicity
and inherent parallel nature of FS-DAM, project the methodology as a
possible effective tool for a second generation High Throughput
Virtual Screening in drug discovery and design.

\section{Acknowledgements} \textcolor{Red}{The computing resources and the related technical
support used for this work have been provided by CRESCO/ENEAGRID High
Performance Computing infrastructure and its staff.\cite{enea}
CRESCO/ENEAGRID High Performance Computing infrastructure is funded by
ENEA, the Italian National Agency for New Technologies, Energy and
Sustainable Economic Development and by Italian and European research
programmes, see http://www.cresco.enea.it/english for information}
\bibliography{ms} 
\bibliographystyle{ms} 

\end{document}